\documentclass[10pt,aps,prd,twocolumn,groupedaddress,showpacs,showkeys]{revtex4}
\usepackage{graphics}
\usepackage{dcolumn}
\newcommand{\be}{\begin{equation}}
\newcommand{\ee}{\end{equation}}
\newcommand{\bea}{\begin{eqnarray}}
\newcommand{\eea}{\end{eqnarray}}
\newcommand{\no}{\nonumber}
\begin{document}

\title{Repulsive 
Casimir forces produced in rectangular cavities: Possible measurements and
applications}

\author{A. Gusso}
\email{gusso@fisica.ufpr.br}

\affiliation{Departamento de F\'{\i}sica, Universidade Federal do Paran\'a,\\
C.P. 19044, 81531-990 Curitiba-PR, Brazil}
\author{A. G. M. Schmidt}
\email{schmidt@uesc.br}
\affiliation{Departamento de Ci\^encias Exatas e Tecnol\'ogicas,
 Universidade Estadual de Santa Cruz, 
CEP 45662-000, Ilh\'eus-BA, Brazil}

\begin{abstract}

We perform a theoretical analysis of a setup intended to  measure the repulsive
(outward)  Casimir forces  predicted to exist inside of  perfectly conducting
rectangular cavities.  We consider the roles of the conductivity of the real
metals, of the temperature and surface roughness. The possible use of this
repulsive force to reduce friction and wear in micro and nanoelectromechanical
systems (MEMS and NEMS) is also considered. 
\end{abstract}



\maketitle


\section{Introduction}
\label{introduction}

Casimir forces are a well known prediction of Quantum Field Theory, and result
whenever the quantum vacuum is subject to constraints. The Casimir forces are
one aspect of a broader subject usually referred to as Casimir effect.
Presently, the Casimir effect finds applications not only in Quantum Field
Theory, but also in Condensed Matter Physics, Atomic and Molecular Physics,
Gravitation and Cosmology, and in Mathematical Physics
\cite{review_bordag,review_greiner,milonni}, and its importance for practical
applications is now becoming more widely appreciated \cite{roukes,importance}.
Not withstanding its importance, the Casimir effect is elusive. The attractive
Casimir forces (ACFs) predicted to exist  between electrically neutral bodies
were measured successfully only a few years ago \cite{review_bordag}.
Presently, the ACF between a sphere (or lens) above a flat disc covered with
metals  is claimed to be measured with an experimental relative error of
approximately 0.27\% at a 95\%  confidence level \cite{mostprecise}, and can be
predicted with a theoretical uncertainty at the level of 1\%. The direct
measurement of the Casimir force between two parallel conducting plates, the
original setup studied by H.~B.~G. Casimir  \cite{original} in 1948, is even
more challenging than for the sphere above a  disc. For that reason it was 
accomplished only recently with a relatively poor precision of $15$\%
\cite{parallel}.

The measurement of such attractive forces sheds some light on the question of
the nature of the quantum electromagnetic vacuum. However, very important
predictions based on the existence of the quantum vacuum have not received the
same attention. This is the case of the repulsive Casimir forces (RCFs).  Such
repulsive forces (outward pressure on the walls) are predicted  to exist inside
of an empty sphere \cite{boyer_sphere} and an empty rectangular cavity  
\cite{analytical,wolfram} with perfectly conducting walls, for the case of
Euclidean space. Such repulsive forces are probably the most striking example
of the geometry dependent nature  of the Casimir effect. However no experiment
was performed to measure RCFs. Only a weak dependence on the geometry was
tested measuring the force between a plate with small sinusoidal corrugations
and a large sphere \cite{nontrivial}. The measurement of the RCF would be one
of the most important probes of the nature of the quantum vacuum  with far
reaching implications.  Because  RCFs have been predicted consistently by
different quantum field theoretic techniques
\cite{boyer_sphere,analytical,wolfram,villareal} if they are  proved not to
exist the physicists will be faced with a new puzzle to be solved. Widely
accepted Casimir energy renormalization and regularization procedures may need
to be reviewed, as suggested by the only dissonant result presented in Ref.
\cite{jaffe}, were no RCFs are found for a rectangular piston.

 The sign of the Casimir force is also predicted to depend upon electric and
magnetic properties of materials. For instance, in Ref.~\cite{boyer_repulsive}
it is anticipated that a repulsive force will exist between two parallel plates
if  one is a perfect conductor and the other  is perfectly permeable. More
recently,  a repulsive force  between two parallel plates  made from dielectric
materials with nontrivial  magnetic susceptibility was anticipated
\cite{kenneth}. However, this effect, which could have interesting applications
for MEMS and NEMS, has not been verified  experimentally and no dielectric
material exists satisfying the requirements  on the values of the magnetic 
susceptibility.

\begin{figure*}
\includegraphics{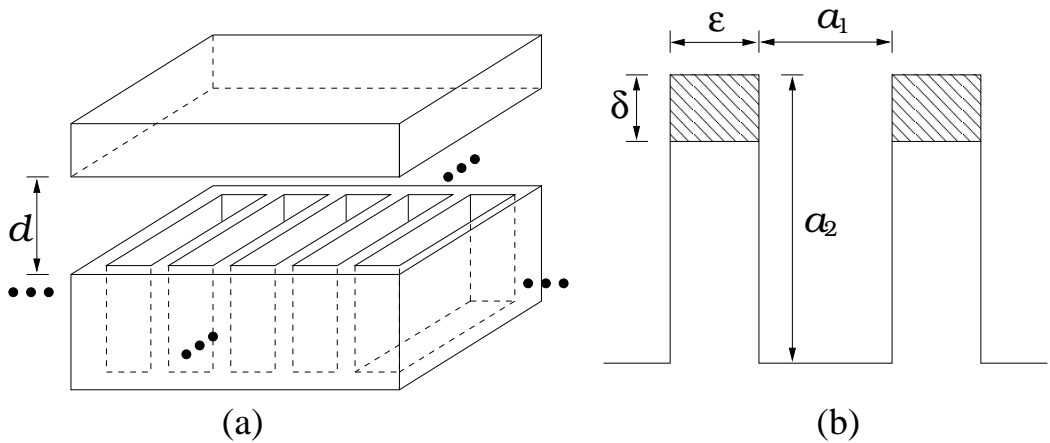}
\caption{\label{setup} (a) A view of the setup including the rectangular
cavities and the plate. The dots denote the possibility of having more
cavities arranged side by side. (b) The definitions of the lengths of the
cavities and walls  shown in a side view.}
\end{figure*}

In spite of the fact that  RCFs are predicted for sphere and rectangular cavity
with perfectly conducting walls, it is reasonable to expect, as for the case of
parallel plates, that they will also be present inside cavities made from good
conductors. For that reason, in this article we address the most important
practical aspects to be taken into account in an experiment intended to measure
the force exerted on one of the walls of a rectangular cavity: the finite
conductivity and roughness of the walls and plate, and the  temperature. The
choice of rectangular cavity instead of the sphere is based primarily on the
fact that  the former could be most easily fabricated with the available
techniques for the fabrication of MEMS and NEMS. We consider the experimental
setup to measure the RCFs to be made of a series of microscopic metallic
rectangular cavities  arranged side by side, forming an array, with one of the
walls open. The repulsive force is then measured by bringing near a plate with
a flat metallic surface.  The forces on the plate are then measured. The use of
a series of cavities, instead of a single cavity, generates  stronger forces
that can be measured most easily. This setup is presented schematically in
Fig.~\ref{setup}(a). We note that a different setup was considered in 
Ref.~\cite{sphere} were a  sphere is used instead of a flat plate. However, it
has to be mentioned that  in Ref.~\cite{sphere} it is considered the case in
which the radius of the sphere is comparable to the cavity length, implying
that the ends of the cavities are left essentially uncovered, and it is not
clearly explained why in this case one still can expect the emergence of
repulsive forces between the cavities and the sphere.  Furthermore, the
roughness of the walls and the role of the temperature are not considered in
the analysis there presented. It is also not explained how the finite 
conductivity of the cavity walls and the sphere were taken into account in the
calculation of the attractive and repulsive Casimir forces. 

 Its worth  to mention that to analyze the flat plate-rectangular cavities
configuration is specially relevant  because the moveable pieces of MEMS and
NEMS typically involve flat surfaces (for example, the rotary pieces in
micromotors and gears), and it is natural to ask whether metallic rectangular
cavities could be used to make such pieces to levitate or, at least, to have
their weight or other undesirable forces partially compensated by a repulsive
force. For that reason, following the analysis on the RCF measurement we
present an analysis on the possible application of repulsive Casimir forces in
MEMS and NEMS to circumvent the problems resulting from friction and wear.
   
\section{Casimir energy and forces}
\label{forces}

In this section we argue that for a setup like that presented in
Fig.~\ref{setup} the resulting Casimir force on the plate is given by the sum
of two independent contributions, namely, the RCF produced by the
electromagnetic vacuum modes inside the cavity and  the attractive force
between the plate and the upper portion of the cavities.

The renormalized Casimir energy inside a rectangular cavity with perfectly
conducting and perfectly smooth walls at zero temperature can be derived in
various manners \cite{review_bordag}. A simple expression suitable for
numerical calculations was derived in  \cite{wolfram}, and in terms of the
internal dimensions  of the cavity $a_1,a_2$ and  $a_3$ it reads
\bea
E_C &=& -\hbar c \frac{a_1 a_2 a_3}{16 \pi^2}
{\sum_{l,m,n=-\infty}^{\infty}} [(a_1 l)^2 +(a_2 m)^2 +(a_3 n)^2 ]^{-2} \no \\
&&+\hbar c \frac{\pi}{48}  \left(  \frac{1}{a_1} +
\frac{1}{a_2}+\frac{1}{a_3} \right).
\label{ec_wolfram}
\eea
The term with $n_1=n_2=n_3=0$ is to be omitted from the summation. From the
principle of virtual work the force on the walls perpendicular to  the
direction of $a_i$ is simply
\be
F_i = - \frac{\partial E_C}{\partial a_i},
\label{force}
\ee
and ranges from positive (outward) to negative (inward) depending on the
relative sizes of $a_1, a_2$ and $a_3$. The Eqs.~(\ref{ec_wolfram}) and 
(\ref{force}) allow one to search for  the configuration of the cavity
resulting into the strongest outward  forces on the walls. The numerical
analysis performed in Ref.~\cite{maclay}, using the above expression for
the energy, suggests  that the forces  $F_2$ and $F_3$ are larger (and
positive) in a configuration satisfying  $a_1 \ll a_2 \ll a_3$, corresponding 
to an elongated parallelepiped. For such a configuration $F_1$ is directed
inward.

Fortunately, whenever  $a_1 \ll a_2 \ll a_3$ for a rectangular cavity we can
use a simple analytical  expression for the Casimir energy
\cite{review_bordag,analytical}
\be
E_C = -\hbar c \left[ \frac{\pi^2 a_2 a_3}{720 \, a_1^3} + 
\frac{\zeta_R (3)}{16 \pi} \frac{a_3}{a_2^2} - \frac{\pi}{48} \left ( 
\frac{1}{a_1} + \frac{1}{a_2} 
\right ) \right],
\label{energy}
\ee
where $\zeta_R$ denotes the Riemann zeta function. The expressions for the two
outward forces are then calculated using Eq.~(\ref{force}) and the result
is
\be
F_2 = \hbar c \left[ \frac{\pi^2 a_3}{720 \, a_1^3} - \frac{\zeta_R (3)}{8 \pi} 
\frac{a_3}{a_2^3} + \frac{\pi}{48 \, a_2^2} \right],
\label{force2}
\ee
and
\be
F_3 =\hbar c \left[ \frac{\pi^2 a_2}{720 \, a_1^3} + \frac{\zeta_R (3)}{16 \pi} 
\frac{1}{a_2^2} \right].
\label{force3}
\ee
These formulas for the forces reproduce the results obtained from 
Eq.~(\ref{ec_wolfram}) to better than $1$,  and because the first term
dominates over the others the forces are positive. 

\begin{figure}
\includegraphics{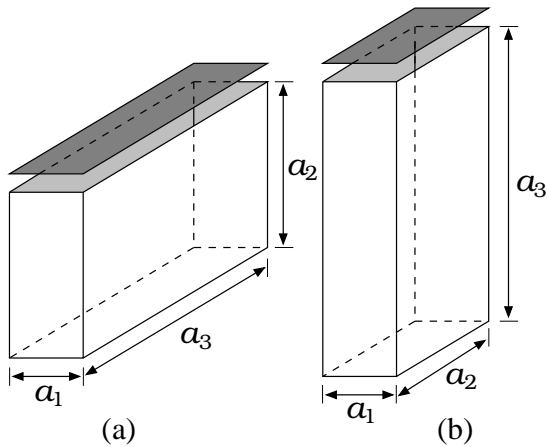}
\caption{(a) Cavity lying horizontally below a plate. (b) Cavity standing
vertically below a plate. Plate is in dark gray and the  open wall is indicated
in light gray.}
\label{orient}
\end{figure}

Therefore there are two possible configurations for our system. In one
configuration  elongated cavities with height $a_2$ are lying horizontally below
the plate, like the cavity in Fig.~\ref{orient}(a), with the force exerted on
the plate corresponding to $F_2$. In the other configuration the cavities are
standing vertically below the plate, like the cavity in Fig.~\ref{orient}(b),
with the force exerted on the plate corresponding to $F_3$. If there where no
other forces acting on the plate,  measuring the RCFs would be a relatively
easy task. The plate could be brought near the open wall closing it completely,
thus  assuring the existence of the electromagnetic vacuum modes that lead to
the Casimir energy Eq.~(\ref{ec_wolfram}). However, when the plate is close to
the top of the walls there will be a resulting attractive force that can
surpass the repulsive force. For that reason, in what follows we analyze which
configuration is the more adequate for an experiment intended to measure RCFs,
delivering the stronger repulsive force compared to the attractive forces
between the plate and the cavity walls.

The repulsive forces exerted on the plate, $F_2$ or $F_3$, are expected to
decrease with increasing $d$, the distance from the plate to the top of the
walls [see Fig.~\ref{setup}(a)]. However, a detailed estimation of this decrease
is beyond the scope of  the present work. Instead we are going to assume in our
calculations that the forces $F_2$ and $F_3$ do not depend on $d$. We expect
this is a reasonable  assumption whenever $d$ is sufficiently small.  In order
not to disturb the electromagnetic vacuum modes that give the most important
contributions to the Casimir energy. This expectation is based on the fact that
in the case the aperture at the top of a cavity is smaller than $\lambda$ the
transmission of the modes to outside the cavity is kept small \cite{apertures}.
Now the values of $\lambda$ that give the most important contribution to the
Casimir energy for a cavity satisfying the condition   $a_1 \ll a_2 \ll a_3$, a
geometry that closely resembles that of two parallel plates (more on that in
Section \ref{corrections}), are those of the order of the smallest edge $a_1$.
Thus, it is reasonable to expect that $F_2$ and $F_3$ are constant up to $d
\alt a_1/2$. For $d \agt a_1/2$ the RCF will certainly decrease, and for that
reason  the relevant distance in an actual experiment is restricted to $d \alt
a_1$.

Because $d\alt a_1$, and $a_1 \ll a_2$ and $a_3$, we can separate the total 
Casimir force between the plate and the cavities into two components: the RCF,
$F_2$ and $F_3$ and the ACF between the plate and the top of the walls. This
conclusion is not as trivial as it may seen to be. If the plate were at
relatively large distances from the cavities the Casimir energy for the 
plate-cavities system  should be calculate from first principles considering
the whole intricate geometry of the cavities. That means, the analysis should
be similar to that carried on for periodically deformed objects in
Ref.~\cite{periodical}. That would also be the case whether $a_1 \sim a_2$,
that means in the case of shallow cavities. For the deep cavities we are going
to consider, only the interaction  between the top of the walls and the plate
is responsible for the attractive force.

Because of the nontrivial geometry involved, in order to calculate  the ACF
between the plate and the cavity walls  we use the  pairwise summation
technique \cite{pairwise1,pairwise2}. This technique was shown to give reliable
results for the Casimir force between  bodies  of arbitrary shape. For
instance, for the force between a flat plate and a small body of arbitrary
shape the maximum possible error was estimated to be 3.8 \% \cite{pairwise1}
when compared to the exact results obtained by quantum field theoretic
techniques.

\begin{figure}
\includegraphics{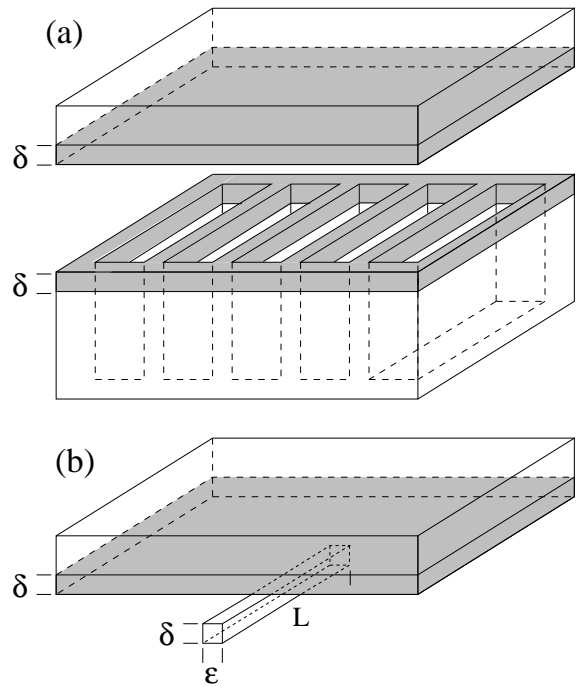}
\caption{\label{top} (a) The relevant volumes of integration highlighted in
light gray. (b) Parallelepiped below a large plate.}
\end{figure}

 In the pairwise summation technique the Casimir energy  is given by
\be
 E_C^{\mathrm{pw}} = -\hbar c \, \Psi(\epsilon_{20}) \int_{V_1} d^3 r_1 
 \int_{V_2} d^3 r_2 \mid
 \mathbf{r}_2 - \mathbf{r}_1 \mid^{-7},
 \label{e_pair}
 \ee
where $V_1$ and $V_2$ are the volumes of the two interacting bodies, and
$\Psi(\epsilon_{20})$  is a constant which depends on the materials on $V_1$
and $V_2$. This expression for the energy does not take into account the fact
that  the pairwise interaction between the atoms in the volumes $V_1$ and $V_2$
are actually screened by the surrounding atoms. In order to partially correct
for this fact avoiding to overestimate the attractive forces  we do not
integrate over the entire volume of the walls and the plate.  Instead we
consider that interactions are only relevant up to a distance $\delta$ inside
the metal. The resulting volumes of integration $V_1$ and $V_2$ are shown in
Fig.~\ref{top}, highlighted in light gray. Clearly, from Eq.~(\ref{e_pair}) it
is not important to define which volume corresponds to $V_1$ and $V_2$ since
the variables are interchangeable. For the constant $\Psi(\epsilon_{20})$ we
take its value in the limit of perfect conductors  $\Psi(\epsilon_{20}) =
\pi/24$, and introduce the corrections due to the finite conductivity later
(Section \ref{corrections}). 

 In order to get results that are independent of the exact number of cavities
in the experimental setup, making the analysis more general, we employed  a
simple strategy. We note that the top of the walls, highlighted in light gray
in Fig.~\ref{top}(a), can be divided into a series of parallelepipeds.
Therefore, because of the additivity of the Casimir energy in the context of
the pairwise summation technique, the final Casimir energy between the plate
and the walls is given by the sum of the individual energy of each segment (a
parallelepiped) and the plate. The only restriction is that the plate must be
sufficiently large to be considered infinite, making the calculations
independent of the actual location of the different segments on the top of the
walls. This condition can be easily fulfilled by a plate only slightly larger
than the array of cavities because the Casimir energy decreases very rapidly
with the increasing distance.

We evaluated   $E_C^{\mathrm{pw}}$ analytically with the help of {\it
Mathematica} \cite{mathematica} for a parallelepiped  with dimensions given by
$\epsilon, \delta$ and  arbitrary length $L$, below a plate with thickness
$\delta$ and lateral dimensions that ensure that the final result is close
enough to that for an infinitely large plate. This arrangement is depicted in
Fig.~\ref{top}(b).  An energy per unity of area is obtained  dividing 
$E_C^{\mathrm{pw}}$ by $\epsilon \times L$. The final Casimir energy, including
all the cavities is then obtained multiplying this energy per unit of area by
the top surface of the walls, a procedure equivalent to summing over the
different segments. Moreover, to make our analysis more general we
assume that there is a sufficiently large number of cavities that  we can
calculate $E_C^{\mathrm{pw}}$ for one cavity and then simply multiply it by the
total number of cavities. In such a case the contribution of the outermost
walls that are partially disregarded are negligible. The area that enters in
the calculation of   $E_C^{\mathrm{pw}}$ is then the effective ``attractive
area" per cavity $S_i = (a_1 + a_i + \epsilon)\epsilon$, where $i=2$ or 3
depending  on whether the cavity is lying horizontally or standing vertically
below the plate, respectively. The area of  the plate under the action of  the
forces $F_{2,3}$ is $A_{2,3}= a_1 \times a_{3,2}$. In what follows we always
calculate the repulsive and attractive Casimir forces for only one cavity, but
the results are actually valid for a large number of cavities as explained
previously.

Similarly to the case of ACFs the RPCs are only appreciable when the dimensions
of the cavity are in the micrometer range. This is the first practical aspect
that has to be considered in any experiment. Presently, a structure  like that
in Fig.~\ref{setup} can be made from metals like gold, nickel, copper, and
aluminum with the smallest features with tens of nanometers, and structures 
like the cavity walls  can be made with  high aspect-ratios
\cite{nanofabrication,graighead}. Consequently,  the small dimensions of the
cavities pose no problem if they are kept above a few tens of  nanometers.

The pressures caused by $F_2$ and $F_3$, whenever $a_1 \ll a_2 \ll a_3$ goes
with $1/a_1^2$, as can be seen from Eqs.~(\ref{force2}) and
(\ref{force3}).  Consequently, the smaller the $a_1$ the bigger the pressure,
which is desirable.  However, the smaller the $a_1$ the smaller the ratio
$A_i/S_i$, for a given $\epsilon$, therefore diminishing the ratio between the
repulsive and the attractive forces.  The lower limit on $a_1$ is  then set by
the lower practical limit on $\epsilon$, because the walls  must be thick
enough to ensure good reflectivity to the electromagnetic modes inside  the
cavity. Such thickness is roughly  determined  by the penetration depth of the
electromagnetic field $\delta_0 = \lambda_p/(2\pi)$, with $\lambda_p$ the
plasma wavelength of the metal. For aluminum(gold) $\lambda_p \approx 107(136)$
nm \cite{lambrecht}, implying $\delta_0 \approx 17(22)$ nm. Now we note that
the intensity of the  incident electromagnetic wave a distance $x$ inside the
metal decreases as $I = I_0 {\rm exp}(-2 x/\delta_0)$. We can ensure almost no
transmission of the electromagnetic waves by making  $x_{\rm max} = \epsilon
\approx 2\times \delta_0$. For that reason, we assume that the smallest
possible thickness is $\epsilon = 30$ nm. For such an $\epsilon$ the smallest
$a_1$ is approximately 100 nm. As we will see next, another reason not to take
$a_1$ smaller than 100 nm is that for cavities made from real  metals the RCFs 
are predicted to  decrease  significantly whenever the smaller side of the
cavity is below approximately 100 nm.   

We now point out that the  configuration of the rectangular cavity that can
lead to the largest ratio between the repulsive and attractive forces on the
plate is that of Fig.~\ref{orient}(a). This is so because in spite of the fact
that both configurations can deliver the same outward pressures, in practice 
the ratio $F_2/S_2$ can be made greater (by one order of magnitude)  than the
ratio $F_3/S_3$.  This results from the fact that  the fabrication of a
vertically standing rectangular cavity   with thin  walls  much higher  than 1
$\mu$m would be very difficult. This implies that in general $a_3 \alt 1 \mu$m.
Because for such a configuration  $a_2 \ll a_3$, the force $F_3$ is highly
constrained [see Eq.~(\ref{force3})] compared to $F_2$, which can be made
arbitrarily large since $a_3$ is not constrained. For that reason, in what
follows we consider only  the case the cavities are lying horizontally below
the plate. This configuration is exactly the one depicted in
Fig.~\ref{setup}(a).

\section{Conductivity, roughness and temperature corrections}
\label{corrections}

The first correction to be taken into account here is that of finite
conductivity, which alters both ACFs and RCFs. To this date finite conductivity
corrections were calculated only  for two simple geometries, namely, for two
plane parallel plates and for a sphere above a disc
\cite{review_bordag,lambrecht}. Such calculations are quite involved, and 
similar calculations for a rectangular cavity  and the interaction between the
top of the walls  and the plate  would be even more demanding. Here, instead of
calculating the corrections from first principles we adopt another strategy and
use the results already obtained for plane parallel plates.

In order to justify this approach we note that the Casimir energy for a rectangular cavity satisfying
$a_1 \ll a_2 \ll a_3$  is approximately that in a region with dimensions   $a_2
\times a_3$ between two infinite parallel plates separated by a distance $a_1$ 
\be 
E_C^0 = -\hbar c  \frac{\pi^2 a_2
a_3}{720 \, a_1^3} , 
\ee 
as can be inferred from Eq.~(\ref{energy}). That means the modes inside the
cavity are approximately the same as those between   parallel plates a distance
$a_1$ apart. Hence it is reasonable to assume that the corrections to the
energy and $F_1$ for the rectangular cavity are adequately described  by those
to the energy and force between parallel metallic plates. In the notation of
Ref.~\cite{lambrecht} we  write, $E_C = \eta_E(a_1) E_C^0$ and  $F_1 =
\eta_F(a_1) F_1^0$, where $E_C^0$ and $F_1^0$ are the energy and force for a
cavity with perfectly conducting walls. The functions $\eta_{E,F}(x)$ are the
correction  factors that range from approximately 1 at large separations to
approximately 0 at the shortest distances. These factors depend upon the
materials on the walls through their frequency dependent  dielectric functions.
In our analysis we modeled the  dielectric functions using the plasma model as
done in Ref.~\cite{lambrecht}. In this context, since $\eta_E$ does not
depend upon $a_2$ the correction to the force $F_2$  is the same as that for
the energy. However, in order to be conservative we assume that the correction 
to $F_2$ is the same  as that for $F_1$ ($\eta_F$ is  slightly smaller than
$\eta_E$). In conclusion, we assume that $F_2 = \eta_F(a_1) F_2^0$, where 
$\eta_F(a_1)$ is plotted in  Fig.~1 of Ref.~\cite{lambrecht} and $F_2^0$
is given by Eq.~(\ref{force2}).  

We have chosen to analyze two rectangular cavities with dimensions that have a
good commitment with the need for strong RCF, to be  approximate by 
parallel plates, and to have reasonable aspect-ratios to meet the requirements
of available fabrication techniques, namely
\be
a_1 = 0.1 \mathrm{\mu m}, \, a_2 = 0.5 \mathrm{\mu m}, \, a_3 = 5 \mathrm{\mu
m}, 
\label{a1100}
\ee
and
\be
a_1 = 0.2 \mathrm{\mu m}, \, a_2 = 1 \mathrm{\mu m}, \, a_3 = 5 \mathrm{\mu m}.
\label{a1200}
\ee
We consider cavities made from aluminum, for its excellent reflectivity in a
wide range of frequencies, and gold, a metal widely employed for the
fabrication of MEMS. For the cavity with $a_1 = 0.1(0.2) \mu$m:  $F_2 =
2.1(0.27)$ pN; the pressure is $P_2 = 4.2(0.27)$ N  m$^{-2}$; for aluminum
$\eta_F(a_1) = 0.50(0.68)$; and for gold $\eta_F(a_1) = 0.44(0.62)$. We stress
that the energy inside the cavity for perfectly conducting walls  differs from
that for parallel plates with area $a_2 \times a_3$ by  just 0.8(3)\%,
therefore justifying our assumptions.

The finite conductivity correction for the ACF was taken to be the same as that
for parallel plates, and the force obtained from the use of Eq.~(\ref{e_pair})
is simply multiplied by $\eta_F(d)$. This is certainly a good approximation
whenever the separation $d$ is small compared to $\epsilon$, because in such a
case the  top of the walls and the plate form a system resembling  two parallel
plates. For $d$ comparable or greater than $\epsilon$ we do not expect that
this approximation fails completely. This expectation relies on the fact that
for larger distances the correction factor  is nearly 1  and vary  at a
relatively slow pace, consequently, it is less important.
Another consequence of the finite conductivity is a rapid decay of
electromagnetic fields inside the metal.  It was for that reason  that we
considered a finite  $\delta$ in the calculation of the ACF.  Based on the
values of $\delta_0$ for aluminum and gold we assume $\delta = 50$ nm.

The second correction to the forces that we consider is that of surface 
roughness. What is relevant here is the stochastic roughness in both the cavity
walls and the plate resulting from the fabrication process. As we did for the
finite conductivity corrections we use the results already derived for the case
of two parallel plates. The corrected energy  inside the cavity  can be
obtained from the expression for the corrected force between two plates in
Ref.~\cite{review_bordag} by simply integrating on the separation,
resulting in
\be
E_C^{\rm roughness} = E_C^0 \left[ 1 + 4 \left( \frac{\delta_{\rm disp}}{a_1} 
\right)^2 + 60 \left( \frac{\delta_{\rm disp}}{a_1}\right)^4 \right],
\label{roughness}
\ee
where $\delta_{\rm disp}$ is the dispersion (roughly the amplitude) of the
stochastic  roughness. In this approximation there is no dependence of the
roughness  correction on $a_2$ and $a_3$. Consequently, the force $F_2$ is
corrected by exactly the same factor as the energy. To keep the corrections
below the 1(5)\% level it is required that $\delta_{\rm disp}/a_1 \alt
0.049(0.10)$. That means for a cavity with $a_1 = 100$ nm that the
imperfections on the  walls can be as large as 5(10) nm. Presently, by means of
electron beam  lithography a precision in the level of 1.3 nm has been obtained
for the fabrication of MEMS and NEMS \cite{lithography}. However, most usual
techniques are not that accurate and a precision at the level of 10 nm is most
likely to be found in an experiment \cite{nanofabrication}.  As a first
approximation the ACF could also receive the same correction expressed in
Eq.~(\ref{roughness}). 

Finally we address the role of temperature. An expression for the Casimir
energy inside a rectangular cavity with finite temperature was derived in
Ref.~\cite{temperature}. It corresponds to adding the following terms
to the energy in Eq.~(\ref{ec_wolfram})
\begin{widetext}
\bea
E_C^{\rm temp} &=& \hbar c \biggl( -\frac{\pi^2 a_1 a_2 a_3}{45 \beta^4} 
 + \frac{\pi}{12 \beta^2}( a_1 + a_2 + a_3 ) \no \\
 &-& \frac{a_1 a_2 a_3}{\pi^2} \sum_{l,m,n,p =1}^{\infty}
\frac{1}{\left[ (a_1 l)^2 + (a_2 m)^2 + (a_3 n)^2 + (\frac{\beta}{2} p)^2 
\right]^2} \no \\ 
&+& \frac{1}{\pi} \sum_{l,p =1}^{\infty} \left\{ \frac{a_1}{\left[ 4(a_1 l)^2 
+ (\beta p)^2 \right]} 
+ \frac{a_2}{\left[ 4(a_2 l)^2 + (\beta p)^2 \right]} 
+  \frac{a_3}{\left[ 4(a_3 l)^2 + (\beta p)^2 \right]} \right\} \biggr),
\eea
\end{widetext}
where $\beta = \hbar c/k_B T$, with $k_B$ the Boltzmann's constant and $T$ the
absolute temperature. For the cavity with $a_1 = 0.1(0.2) \mu$m at a
temperature of 300 K the energy decreases significantly by  1.1(4.7) \%, while
the force $F_2$  decreases just 0.08(0.2)\%. The corrections are still very
small at higher temperatures.

\section{Force measurement}
\label{measurement}

In the present  analysis of force measurement we disregard both the temperature
and roughness corrections. The former because the correction to the force is
always much smaller than the expected experimental accuracy on the force
measurement; of the order of a few percent in any realistic scenario. The
later  because it could be made suitably small depending on the fabrication
technique. Yet, we consider the most important correction, that of the finite
conductivity, that reduce the repulsive force produced by the cavity by half of
its original value and the attractive forces by even greater factors
\cite{lambrecht}. Besides being the most important correction, greatly
exceeding the expected corrections due to roughness, the conductivity will
depend  upon the material the cavity and  the plate are made from and only
marginally on the fabrication technique. In this sense, the corrections due to
conductivity are universal, and do not depend upon the specific fabrication
technique that will be employed, henceforth justifying the present theoretical
analysis. 

As already mentioned in Section \ref{forces} in the analysis we did not model
the expected decrease on the repulsive force as a function of the separation
$d$.  However, it is  reasonable to assume that for small separations the
repulsive force can be well described by Eq.~(\ref{force2}). For small
separations we mean $d$ small compared to $a_1$, because it is the smaller
cavity dimension that, after all, determines the the smaller frequencies
allowed inside the cavity.  For small $d$ we can expect a small perturbation on
the modes inside the cavity in a large range of frequencies, henceforth
ensuring the existence of the repulsive force.

\begin{figure}
\begin{center}
\includegraphics{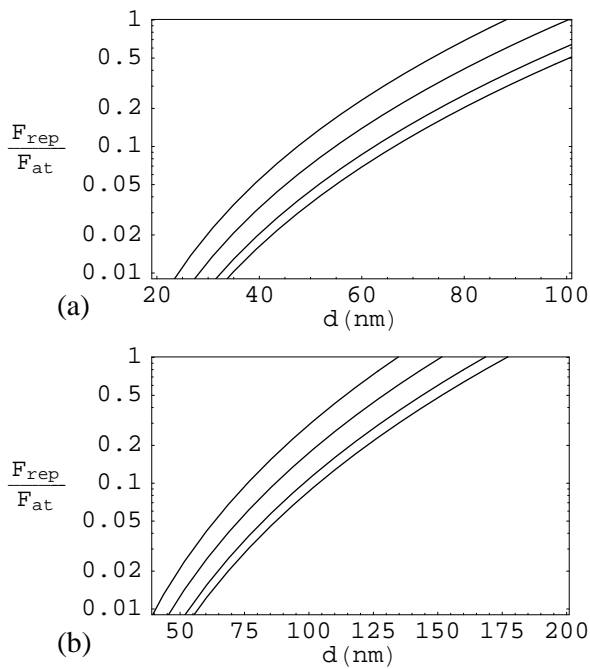}
\end{center}
\caption{\label{ratio} The ratio between repulsive ($F_{\rm rep}$) and 
attractive ($F_{\rm at}$) Casimir forces for a rectangular cavity as a function
of the separation $d$. From the upper to the lower curve  $\epsilon =$30, 50,
80, and  100 nm. Cavity dimensions in (a) given by Eq.~(\ref{a1100}) and 
in (b) by Eq.~(\ref{a1200}). }
\end{figure}

The most important information for an experiment designed to measure RCFs is
the ratio between the repulsive and attractive forces as a function of the 
separation $d$. In Figs.~\ref{ratio}(a) and \ref{ratio}(b) we present exactly 
this ratio for the cavities with $a_1=0.1 \mu$m and $0.2 \mu$m, respectively.
The results are for the cavity and the plate made from gold, however
essentially the same results are obtained for aluminum. We considered four
different $\epsilon$, from the smallest possible value to one that could be
most easily obtained by the presently available fabrication techniques. 

\begin{figure}
\begin{center}
\includegraphics{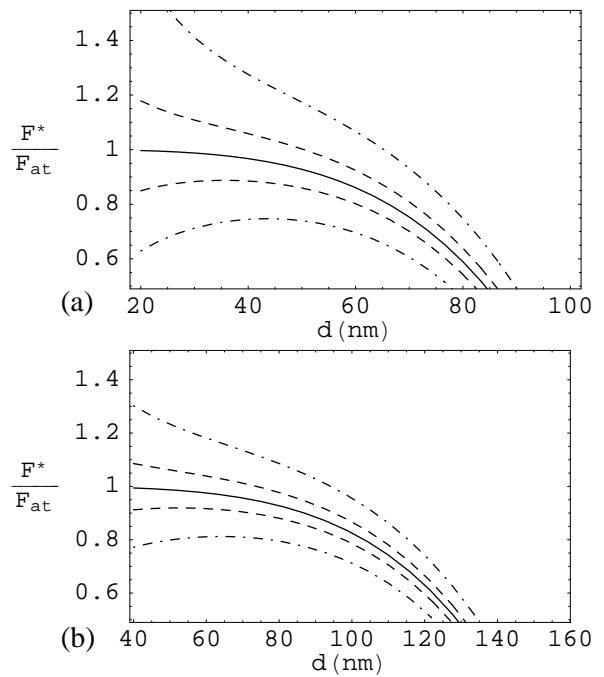}
\end{center}
\caption{\label{errors} Curves for the ratio defined in Eq.~(\ref{errorfrac})
for $\Delta = 0$  (continuous), $\Delta = \pm 1$ nm (dashed) and $\Delta =  \pm
3$ nm  (dot-dashed).Cavity dimensions in (a) given by Eq.~(\ref{a1100}) and in
(b)  by Eq.~(\ref{a1200}). }
\end{figure}

What we can infer from figure~\ref{ratio} is the smallness of the repulsive
force compared to the attractive one in the range of distances at which our
calculations are  more reliable and precise ($d \alt a_1/2$). The ratios are
larger for the cavity with $a_1 = 200$ nm, and for $d = a_1/2 = 100$ nm the RCF
amounts to 30\% of the ACF. For the cavity with $a_1 = 100$ nm the RCF amounts
to only 10\% at $d=a_1/2$. As a consequence of the smallness of the ratios
$F_{\rm rep}/F_{\rm at}$, any measurement of the force exerted on the top
plate  has to be very precise. For the static measurement of the force on the
top plate a precise knowledge of the separation $d$ is  also required. This
fact can be illustrated  by  the ratio between  the sum of the attractive force
at the actual position  and the repulsive force  and the attractive force at
the distance $d$ as determined from the experiment,  
 \be
 \frac{F^*}{F_{\rm at}} =\frac{F_{\rm at}(d+\Delta) + 
 F_{\rm rep}}{F_{\rm at}(d)} , 
 \label{errorfrac}
 \ee
 where $\Delta$ represents the relative displacement to the measured distance
 due to the uncertainties. The curves for this ratio are presented in
 Fig.~\ref{errors} for the two cavities and for $\Delta = 0,\pm 1$ and $\pm 3$
 nm. The upper(lower) curves are for negative(positive) $\Delta$. We note that
 even for $\Delta = \pm 1$ nm the errors are in the range $5-15$\% and are of
 the order of the force the experiment intends to measure (see
 Fig.~\ref{ratio}). Consequently, the distance has to be measured with an
 accuracy  better than 1 nm. In order to estimate the required accuracy we note
 that for a nominal separation $d=50$ nm, an inaccuracy of $0.2$ nm implies an 
 uncertainty in the force measurement of approximately $\pm 1.5$\% for both
 cavities, which is acceptable.

 Our analysis leads to the conclusion that very stringent requirements have to
 be satisfied by the experimental setup in order to allow for an adequate
 measurement of the RCF in a rectangular cavity. Such requirements surpass
 considerably those for the experiments already carried out for the
 measurement of ACFs \cite{review_bordag,mostprecise,parallel}.

 \section{Applications}
 \label{applications}
 
 As already mentioned in section \ref{introduction}, repulsive forces could
 have interesting applications in MEMS and NEMS. In fact, such  forces could be
 the solution for  the problems that are presently imposing severe
 restrictions  on the functioning of MEMS with moveable parts, namely, 
 friction and wear \cite{failure}. The forces caused by friction are usually
 very large at small scales \cite{wang} when compared to the forces that can be
 delivered by the available driven systems in, e.g., micromotors and
 microactuators. Usually friction obeys Amonton's law  (frictional force
 depends linearly on the load through the coefficient of friction), however, at
 small scales friction turns out to be proportional to the contact area between
 the surfaces \cite{wang}. For systems sufficiently large to obey Amonton's
 law, repulsive forces could be used to reduce the load. For instance, the
 rotary piece of a micromotor or gear (usually with the shape of a disc) could
 be lifted by a bottom force that could partially or completely compensates for
 its weight. This force could be the RCF predicted in Ref.~\cite{kenneth}
 or, as we propose here, the force produced by a set of rectangular cavities
 placed beneath the rotary piece of the micromotor or gears. The  first option
 requires the use of suitable materials that presently are not available, and
 is still a matter of debate whether such forces could actually exist
 \cite{comment,henkel}. The second option, the use of cavities beneath the
 moveable pieces, could be a simple solution whenever these pieces  were made
 from metals or could at least be covered with a thin metal layer.  For smaller
 systems, where the load does not play the most important role in the resulting
 frictional force, the repulsive forces could be used in the same way to reduce
 the effective weight that has to be sustained by the  rotating pivots or
 bearings. Consequently, the pivots and bearings could possibly be smaller,
 leading to a reduction in the frictional  force and wear. Such a reduction is
 highly desirable since wear is the most important source of failure in MEMS ,
 limiting their continuous operation lifetime to be of the order of seconds or
 minutes rather than hours or days \cite{failure}.

 To estimate the capability of the repulsive force produced by the rectangular
cavities to compensate for the weight of the moveable parts of MEMS and NEMS we
determined the distance $d_0$ at which the RCF equates the ACF and the
distances required to the repulsive force to equate the ACF added to the weight
of a plate made from a metal with an intermediate density $\rho = 8.9$ g
cm$^{-3}$ (similar  to that of nickel and cooper) and thickness of 1 $\mu$m
and  10 $\mu$m denoted $d_1$ and $d_{10}$, respectively. At this point we have
to note that  structures with thickness ranging from 0.1 $\mu$m up to 10 $\mu$m
are usually employed in the fabrication of parts of MEMS and NEMS
\cite{graighead} even  when the other dimensions of these parts are of the
order of a few millimeters \cite{sandia}. We present in Table  \ref{table1}
$d_0,d_1$ and $d_{10}$ for the cavity with $a_1=0.1 \mu$m, made  from aluminum
and gold, and  for the thickness of the walls $\epsilon = 30$ nm and 50 nm. In
Table \ref{table2} the results are presented for the cavity with  $a_1=0.2
\mu$m. In this case, because the repulsive force is not strong  enough to
equate the weight of a plate 10 $\mu$m thick, we present the distance $d_{0.1}$
required to sustain a plate with thickness of 0.1 $\mu$m.

 \begin{table}
 \caption{\label{table1} The distances $d_0,d_1$ and $d_{10}$ as defined on
 the text for cavities made from Al(Au), and dimensions given in Eq.~(\ref{a1100}).}
 \begin{ruledtabular}
 \begin{tabular}{cccc}
 $\epsilon$ (nm) & $d_0$ (nm)& $d_1$ (nm)& $d_{10}$ (nm) \\ \hline
 30 & 88.2(88.1) & 89.4(89.5) & 107(112) \\
 50 & 100(100) & 102(102) & 126(135) \\
 \end{tabular}
 \end{ruledtabular}
 \end{table}

 In order to better understand the implications of the results presented in
Tables \ref{table1} and \ref{table2} we have to remember the fact that the RCF
produced by the cavity on the plate is expected to decrease with the
separation $d$. Actually, from simple wave propagation arguments,  the change
on the force is expected to depend upon the ratio $d/a_1$.  Consequently, we
can expect a smaller correction (smaller decrease) to the force due to the
separation  for the cavity for which the ratio $d/a_1$ is smaller. We now note
that for $\epsilon = 30$ nm $d_0(d_1)$ corresponds to 88(89)\% and 68(82)\% of
the cavity width for $a_1 = 0.1 \mu$m and 0.2$\mu$m, respectively.  For that
reason  the cavity with $a_1 = 0.2 \mu$m is the most adequate for
investigations concerning the reduction of friction and wear.  Because
$d_{0.1}$ is only slightly larger than $d_0$, levitation of thin metallic
plates caused by RCF is also likely to occur. 

\begin{table}
 \caption{\label{table2} The distances $d_0,d_{0.1}$ and $d_1$ as defined on
 the text for cavities made from Al(Au), and dimensions given in Eq.~(\ref{a1200}).  }
 \begin{ruledtabular}
 \begin{tabular}{cccc}
 $\epsilon$ (nm) & $d_0$ (nm)& $d_{0.1}$ (nm)& $d_1$ (nm) \\ \hline
 30 & 135(134) & 137(136) & 163(166) \\
 50 & 152(151) & 154(153) & 186(192) \\
 \end{tabular}
 \end{ruledtabular}
 \end{table}   
 
We also suggest the use of an artifice in order to reduce further the distances
at which repulsive forces could counterbalance the attractive forces: thiner
walls with short height built  on  top of the cavity walls. Thiner walls may
assure enough reflectivity for the electromagnetic modes inside the cavity
without further disturbing  the modes if they are kept sufficiently short. The
small aspect-ratio further facilitates their fabrication. For instance,  if the
top  walls were 15 nm thick and 45 nm high the distances between the top of
these walls and the plate are predicted to be those presented in Tables
\ref{table3} and \ref{table4}.  In calculating those distances we summed over
the contributions from the original wall and the additional top wall.  The
contribution of the original wall is small as can be seen from the similarity
between the results for $\epsilon = 30$ nm and 50 nm in Tables  \ref{table3}
and \ref{table4} as compared to the results in Tables \ref{table1} and
\ref{table2} that differ considerably.        

 \begin{table}
 \caption{\label{table3} The distances $d_0,d_1$ and $d_{10}$ as defined on
 the text for cavities made from Al(Au), with the dimensions given in Eq.~(\ref{a1100}), and with additional top walls.}
 \begin{ruledtabular}
 \begin{tabular}{cccc}
 $\epsilon$ (nm) & $d_0$ (nm)& $d_1$ (nm)& $d_{10}$ (nm) \\ \hline
 30 & 78.4(78.3) & 79.7(79.7) & 96.6(102) \\
 50 & 82.0(82.0) & 83.5(83.8) & 107(116) \\
 \end{tabular}
 \end{ruledtabular}
 \end{table}    

It is clear that the introduction of the top walls can considerably reduce the
required separations. If the top walls can be made thiner and taller without
further disturbing the modes inside the cavity is a subject that deserves
further theoretical and experimental investigation. Triangular structures are
also worth of investigation. Anyhow, for the shorter distances thus obtained
the assumption of a constant RCF as $d$ varies is more reliable, and therefore
the results are self-consistent.

\section{Final discussion and conclusions}
\label{conclusion}

In this article we presented  a realistic analysis of a setup
intended to measure the repulsive forces  resulting from the geometrical
constraints imposed on the quantum electromagnetic vacuum. For realistic we
mean that the nonideality of the cavity was taken into account in the
calculation of the RCF as well as the unavoidable ACF. We took advantage of the
similarity between a rectangular cavity satisfying the condition  $a_1 \ll a_2
\ll a_3$ and two plane parallel plates, considerably simplifying the analysis.
The results thus obtained are expected to be a very good description of the 
reality for small ratios $d/a_1$ and still reliable when the ratio is around
$0.5$.   

\begin{table}
 \caption{\label{table4} The distances $d_0,d_{0.1}$ and $d_1$ as defined on
 the text for cavities made from Al(Au), with the dimensions given in Eq.~(\ref{a1200}), and with additional top walls.}
 \begin{ruledtabular}
 \begin{tabular}{cccc}
 $\epsilon$ (nm) & $d_0$ (nm)& $d_{0.1}$ (nm)& $d_1$ (nm) \\ \hline
 30 & 125(124) & 127(126) & 153(157) \\
 50 & 132(132) & 134(134) & 166(173) \\
 \end{tabular}
 \end{ruledtabular}
 \end{table}

From the results presented in section \ref{measurement} we conclude that for the
smaller separations at which our approach is more precise, attractive forces
are always considerably greater than the attainable repulsive forces. This fact
poses severe requirements for the experiment.  For separations larger than
approximately $a_1/2$ a reduction of the repulsive force is  expected  and the
curves in Figs.~\ref{ratio} and \ref{errors} are no longer precise. However,
these curves indicate that even under the more optimistic assumption that the
decrease in the RCF is small and that the reliability of our results extends to
larger separations, the measurement may be difficult, unless the cavity walls
are sufficiently thin. That this  is specially true for the case the cavity has
$a_1 = 0.1 \mu$m, can be seen from the fact that  for $\epsilon = 100$ nm, at a
separation $d=a_1=0.1 \mu$m the repulsive force amounts to only 50\% of the
attractive force. Fortunately, we have a better situation for the case of a
larger cavity  since the RCF equates the ACF at  shorter distances, as
can be seen in Fig.~\ref{ratio}(b). 

It is worth to mention that there seems to be no advantage on the use of 
cavities with $a_1$ much greater than 200 nm. The reason for that is the fact
that  for cavities with larger $a_1$ the ratio $F_{\rm rep}/F_{\rm at}$ is
essentially the same as that for $a_1= 200$ nm when plotted as a function of
$d/a_1$. Nevertheless, the RCF and ACF decrease significantly, possibly making
its measurement less precise. This fact has to be considered in the design of
any actual experiment.

The use of the plasma model in the calculation of the finite conductivity
corrections results in correction factors $\eta_F$ that are from 2\% to 10\%
smaller  than those predicted using the tabulated data for the dielectric
functions of aluminum and gold for distances around 100 nm \cite{lambrecht}.
This fact along with our conservative assumption that the force $F_2$ is 
corrected by the factor $\eta_F$ for the force $F_1$, may imply that the actual
repulsive forces delivered by the cavities in an experiment are greater than the
ones we predicted here by at most 20\%. Such an increase in the force does not
significantly changes our results because of the strong dependence of the
attractive forces on the separation $d$. More precisely, the distances would 
decrease no more than 5\%.

The most obvious use of the RCF in MEMS and NEMS is to levitate structures as
we proposed here, preventing friction and wear. However, the  applicability of
such forces is conditioned   to the actual decrease of the RCF with the
separation between the cavities and the upper (plate-like) structure. As
already mentioned the determination of the actual repulsive force with the
distance is beyond the scope of the present work, and is expected to be quite
involved, specially in the case that the finite conductivity of the walls are
taken into account. Nonetheless, the results presented in section
\ref{applications}, based on the extrapolation of a constant RCF to larger
separations, indicate that  the RCF produced by the rectangular cavity is
potentially useful and the importance  of the reduction of wear and friction in
MEMS and NEMS makes it worth of further investigation.

\begin{acknowledgments}

The authors gratefully acknowledge CNPq for research fellowships.
A.~G.~M.~Schmidt was also partially supported by Funda\c{c}\~ao de Amparo \`a
Pesquisa do Estado da Bahia (FAPESB). A.~Gusso.~ is thankful to
I.~A.~H\"ummelgen, D.~H.~Mosca and E.~S.~Silveira for useful conversations.

\end{acknowledgments}


\begin{thebibliography}{99}

\bibitem{review_bordag} M.~Bordag, U.~Mohideen, and V.~M.~Mostepanenko, 
Phys.~Rep.~{\bf 353}, 1 (2001).

\bibitem{review_greiner} G.~Plunien, B.~M\"uller, and W.~Greiner, Phys.~Rep.~
{\bf 134}, 87 (1986).

\bibitem{milonni} P.~W.~Milonni, \textit{The Quantum vacuum: An Introduction to
Quantum Electrodynamics}, (Academic Press, New York, 1994).

\bibitem{roukes} E.~Buks and M.~L.~Roukes, Nature {\bf 419}, 119 (2002).

\bibitem{importance} H.~J.~de los Santos, Proc.~IEEE {\bf 91}, 1907 (2003);
G.~J.~Maclay, H.~Fearn, and P.~W.~Milonni, Eur.~J.~Phys.~{\bf 22}, 463 (2001).


\bibitem{mostprecise} R.~S.~Decca {\it et al.}, Phys.~Rev. D {\bf 68}, 116003
(2003).

\bibitem{original} H.~B.~G.~Casimir, Proc.~K.~Ned.~Akad.~Wet.~ {\bf 51}, 793
(1948).

\bibitem{parallel} G.~Bressi, G.~Carugno, R.~Onofrio, and G.~Ruoso, Phys.~ 
Rev.~ Lett.~ {\bf 88}, 041804 (2002).

\bibitem{boyer_sphere} T.~H.~Boyer, Phys.~Rev.~ {\bf 174}, 1764 (1968).

\bibitem{analytical} S.~G.~Mamaev and N.~N.~Trunov, Theor.~Math.~Phys.~(USA)
{\bf 38}, 228 (1979); Sov.~Phys.~J.~ {\bf 22}, 966 (1979).

\bibitem{wolfram} J.~Ambj{\o}rn and S.~Wolfram, Ann.~Phys.~(N.~Y.~) {\bf 147}, 1
(1983). 

\bibitem{nontrivial} Anushree Roy and U.~Mohideen, Phys.~Rev.~Lett.~ {\bf 82},
4380 (1999).

\bibitem{villareal} S.~Hacyan, R.~J\'auregui, and C.~Villarreal,  
Phys.~Rev.~A {\bf 47}, 4204 (1993).

\bibitem{jaffe} M.~P.~Hertzberg, R.~L.~Jaffe, M.~Kardar, and A.~Schardicchio,
quant-ph/0509071. 

\bibitem{boyer_repulsive} T.~H.~Boyer, Phys.~Rev.~A  {\bf 9}, 2078 (1974);
F.~C.~Santos, A.~Ten\'orio, and A.~C.~Tort, Phys.~Rev.~D {\bf 60}, 105022
(1999).


\bibitem{kenneth} O.~Kenneth, I.~Klich, A.~Mann, and M.~Revzen,
Phys.~Rev.~Lett.~ {\bf 89}, 033001 (2002).

\bibitem{sphere} J.~Maclay {\it et al.}, published as 
{\it AIAA/ASME/SAE/ASEE 37th
Joint Propulsion Conference}, Salt Lake City, July 8, 2001 (available at 
\url{http://www.quantumfields.com});
G.~J.~Maclay and J.~Hammer, in  {\it Proc. of
the 7th International Conference on Squeezed States and
Uncertainty Relations} edited by D.~Han, Y.~S.~Kim, B.~E.~A.~Saleh, 
A.~V.~Sergienko, and M.~C.~Teich, available only in electronic format at 
\url{http://www.physics.umd.edu/rgroups/ep/yskim/boston/boston.html}

\bibitem{maclay} G.~J.~Maclay, Phys.~Rev.~A {\bf 61}, 052110 (2000).

\bibitem{apertures} P.~M.~Morse and P.~J.~Rubenstein, Phys.~Rev.~{\bf 54},
895 (1938); Min Li {\it et al.}, IEEE Trans.~Eletromag.~Compat.~{\bf 39},
225 (1997)

\bibitem{periodical} T.~Emig, Europhys.~Lett.~, {\bf 62}, 466 (2003); Phys. Rev.
A {\bf 67}, 022114 (2003).  

\bibitem{pairwise1} V.~ M.~ Mostepanenko and I.~Yu.~Sokolov, 
Sov.~Phys.~Dokl.~(USA) {\bf 33}, 140 (1988).

\bibitem{pairwise2} M.~Bordag, G.~L.~Klimchitskaya, and V.~M.~Mostepanenko, 
Mod.~Phys.~Lett.~{\bf 19}, 2515 (1994); Int.~J.~Mod.~Phys.~A {\bf 10}, 2661 (1995).
 

\bibitem{mathematica} S.~Wolfram, \textit{The Mathematica Book}, 4th ed.,
(Wolfram Media/Cambridge University Press, 1999) 

\bibitem{nanofabrication}  C.~R.~K.~Marrian and D.~M.~Tennant,  
J.~Vac.~Sci.~Technol.~A {\bf 21}, S207 (2003).

\bibitem{graighead} H.~G.~Graighead, Science {\bf 290}, 1532 (2000).

\bibitem{temperature} F.~C.~Santos and A.~C.~Tort, Phys.~Lett.~ B {\bf 482}, 323
(2000).
 
\bibitem{lambrecht} A.~Lambrecht and S.~Reynaud, Eur.~Phys.~J.~ D {\bf 8}, 309
(2000).

\bibitem{lithography} J.~T.~Hastings, Feng Zhang, and Henry I.~Smith, 
J.~Vac.~Sci.~Technol.~ B {\bf 21}, 2650 (2003).


\bibitem{failure} S.~L.~Miller {\it et al.}, Microelectron.~Reliab.~ {\bf 39}, 
1229 (1999); J.~A.~Williams, Wear {\bf 251}, 965 (2001); 
W.~Merlijn van Spengen, Microelectron.~Reliab.~ {\bf 43}, 1049
(2003).

\bibitem{wang} Weiyuan Wang {\it et al.}, Sens.~Actuators A {\bf 97-98}, 486 
(2002).

\bibitem{comment} D.~Iannuzzi and F.~Capasso, Phys.~Rev.~Lett. {\bf 91}, 
029101 (2003).
 
\bibitem{henkel} C.~Henkel and K.~Joulain, quant-ph/0407153.

\bibitem{sandia} See available information at 
\url{http://mems.sandia.gov/scripts/index.asp}.


\end{thebibliography}
\end{document}